\documentstyle[11pt,paspconf]{article}

\begin{document}

\title{
Galaxy counts at 450\,$\mu$m and  
850\,$\mu$m
}

\author{A. W. Blain}
\affil{Cavendish Laboratory, Madingley Road, Cambridge, CB3 0HE, UK
}

\author{R. J. Ivison} 
\affil{Department of Physics \& Astronomy, University College London, 
Gower Street, London, WC1E 6BT, UK}

\author{J.-P. Kneib} 
\affil{Observatoire Midi-Pyr\'en\'ees, 14 Avenue E. Belin, 
31400 Toulouse, France} 

\author{Ian Smail} 
\affil{Department of Physics, University of Durham, South Road, 
Durham, DH1 3LE, UK} 




\begin{abstract}
Surveys of the distant Universe have been made using the SCUBA  
submillimeter(submm)-wave camera at the JCMT. 450- and 
850-$\mu$m data is taken simultaneously by SCUBA in the same 
5-arcmin$^2$ field. Deep 850-$\mu$m counts of high-redshift dusty 
galaxies have been published; however, at 450\,$\mu$m both the
atmospheric transmission and antenna efficiency are lower, and the 
atmospheric noise is higher, and so only upper limits to the 
450-$\mu$m counts have been reported so far. Here we apply the 
methods used by Blain et al.\ (1999) to derive deep 
850-$\mu$m counts from SCUBA images of lensing clusters to the 
450-$\mu$m images that were obtained in parallel, in which 
four sources were detected. We present the first 450-$\mu$m galaxy 
count. This analysis has only just become possible because 
the volume of data and the difficulty of calibration are both 
greater for the 450-$\mu$m array.  
In light of recent work, in which the 
identification of two of the galaxies in our sample was clarified, 
we also update our deep 850-$\mu$m counts. 
\end{abstract}


\keywords{galaxies: formation, galaxies: evolution, cosmology: observations, 
gravitational lensing, galaxies: general}


\section{Introduction}

Submm-wave surveys are sensitive to high-redshift dusty galaxies. 
By exploiting the gravitational lensing effect  
of rich foreground clusters, the efficiency of these surveys is  
enhanced as compared with those made in blank fields (Blain 1998). In 
addition, follow-up observations are significantly easier in the 
lensed fields as compared with true blank fields because of the typical 
magnification, a factor of 2.5, of the detected sources at all 
wavelengths 
(Ivison et al.\ 1998, 1999; Frayer et al.\ 
1998, 1999; Smail et al.\ 1998, 1999a,b). 

Surveys using the 850/450-$\mu$m SCUBA (Holland et al.\ 1999) 
have provided deep 850-$\mu$m galaxy counts and upper limits to 
the 450-$\mu$m counts (Smail, Ivison \& Blain 1997; Barger et al.\ 
1998,1999a; Hughes et al.\ 1998; Blain et al.\ 1999; Eales et al.\ 1999).  
Here we use the detections of four high-redshift dusty galaxies in 
the 450-$\mu$m SCUBA lens survey data (Smail et al.\ 1998) to 
yield the first galaxy count at a wavelength of 450\,$\mu$m.

The relative number 
counts at different wavelengths depend on the distribution 
of both the redshifts and dust temperatures of the submm 
galaxy population.
It is possible to impose more rigorous constraints on the form of 
evolution of high-redshift dusty galaxies if accurate galaxy counts are 
available at several submm wavelengths. 

\section{Obtaining counts at 450\,$\mu$m} 

Recently we published counts of galaxies detected at a wavelength 
of 850\,$\mu$m through the cores of seven massive cluster 
lenses 
(Smail et al.\ 1998) to a depth of 
0.5\,mJy (Blain et al.\ 1999). We  
used accurate mass models 
of the foreground lenses, which are constrained using the properties 
of lensed arcs detected in deep optical images and the 
spectroscopic redshifts of multiply-imaged background
galaxies (for example Kneib et al.\ 1993; B\'ezecourt et al.\ 1999),
to reconstruct the background sky. The robustness of the 
method was verified by extensive Monte-Carlo simulations. 
Here we apply the same method to the 450-$\mu$m SCUBA maps 
of the seven lensing clusters observed in the survey; Cl0024+16, 
A370, MS0440+02, Cl0939+47/A851, A1835, A2390 and Cl2244-02. 
The data was taken in a range of (generally exceptional) 
atmospheric conditions, and so the thresholds for the  
detection of a 450-$\mu$m source vary from cluster to cluster. 
The $3\sigma$ flux density limits for detection are about 
60, 30, 60, 20, 20, 60 and 60\,mJy in each cluster respectively.  
Four sources were detected: one behind A370 (SMM\,J02399$-$0136, the 
brightest source in the sample, with a 450-$\mu$m flux density of 
$85 \pm 15$\,mJy); two behind A1835 (SMM\,J14009+0252 and 
SMM\,J14011+0252; Ivison et al.\ 1999); and one behind Cl0939+47/A851 
(an extremely red object or ERO -- SMM\,J09429+4658; Smail et al.\ 
1999a). The 450-$\mu$m counts that result from the analysis, 
which is described in detail in Blain et al.\ (1999), are listed in 
Table\,1 and Fig.\,1. The 450-$\mu$m count at about 10\,mJy is 
equivalent to the 850-$\mu$m surface density at 3\,mJy. 

\begin{table}
\caption{
New 450-$\mu$m and updated 850-$\mu$m integral counts of galaxies. 
Our previous direct 850-$\mu$m counts are listed for comparison. Our 
Monte-Carlo method (Blain et al.\ 1999) yields a 450-$\mu$m 
count of the form $N(>S) = K (S/S_0)^\alpha$ as a function of flux 
density $S$, with $K = 530 \pm 300$\,deg$^{-2}$, $\alpha = -1.8 \pm 0.5$ 
and $S_0 = 20$\,mJy.
} \label{tbl-1}
\begin{center}\scriptsize
\begin{tabular}{cccc} 
Wavelength & Flux density & Count &
Previous count \\
($\mu$m) & (mJy) & (10$^3$\,deg$^{-2}$) & (10$^3$\,deg$^{-2}$) \\
\tableline
\noalign{\smallskip}
450 & 10.0 & $2.1 \pm 1.2$ & ...\\
    & 25.0 & $0.5 \pm 0.5$ & ...\\
\noalign{\smallskip}
850 & 0.25 & $51 \pm 19$ & ... \\
    & 0.5 & $27 \pm 9$ & $22 \pm 9$ \\
    & 1.0 & $9.5 \pm 3.3$ & $7.9 \pm 3.0$ \\
    & 2.0 & $2.9 \pm 1.0$ & $2.6 \pm 1.0$ \\
    & 4.0 & $1.6 \pm 0.7$ & $1.5 \pm 0.7$ \\
    & 8.0 & $0.92 \pm 0.53$ & $0.8 \pm 0.6$ \\
    & 16.0 & $0.34 \pm 0.34$ & ... \\
\end{tabular}
\vskip -0.5cm
\end{center}
\end{table}

\section{Updating the 850-$\mu$m counts} 

In our earlier analysis of the 17 sources detected in the SCUBA lens 
survey to yield 850-$\mu$m counts, we first removed 2 submm 
sources identified with cluster cD galaxies from the sample 
(Edge et al.\ 1999). We also removed one other submm galaxy 
from the sample, which was identified with a spiral galaxy in the 
foreground of 
the cluster Cl0939+47/A851, and identified one further submm 
galaxy with a spiral galaxy falling into MS0440+02. We 
now know that these two SCUBA detections are more likely to be 
identified with EROs discovered in our deep near-infrared follow-up
images (Smail et al.\ 1999a) than with the low-redshift spiral galaxies. 
Here we repeat the earlier analysis of the counts (Blain et al.\ 1999), 
but now include these two galaxies as lensed high-redshift 
background sources. The updated 850-$\mu$m counts that result are 
shown in Table\,1 and Fig.\,1. The new results are within the 1$\sigma$ 
errors of our previous analysis of the 850-$\mu$m counts, and are 
modified at only the 10\% level if the redshift distribution of the 
source population is assumed to be given by the results of either 
Barger et al.\ (1999b) or Smail et al.\ (1999b). 

We emphasize that three of the 850-$\mu$m galaxies  
used in this analysis have flux densities less than 3\,mJy, after 
correcting for lensing magnification. Further, two of these sources 
have flux densities less than the blank-field confusion limit for 
identification (about 2\,mJy), and one has a sub-mJy flux density. 

The lower limits to the background radiation intensity ($\nu I_\nu$) 
obtained from the flux densities 
of the detected galaxies 
are $(5.0 \pm 1.5) \times 10^{-10}$ and 
$(1.1 \pm 0.6) \times 10^{-9}$\,W\,m$^{-2}$\,sr$^{-1}$ 
at 850 and 450\,$\mu$m respectively, 
94\% and 34\% of the {\it COBE}-FIRAS values 
(Fixsen et al. 1998).  

\section{Conclusions} 

We present the first count of galaxies at the short submm wavelength 
of 450\,$\mu$m, and update our deep 850-$\mu$m counts in the light 
of improved optical identifications (Smail et al.\ 1999a). The relative 
counts of galaxies at 450 and 850\,$\mu$m are consistent with both 
the spectral energy distributions and the forms of evolution of distant 
dusty galaxies that were derived and discussed in our modeling 
papers. 

\begin{figure}[t]
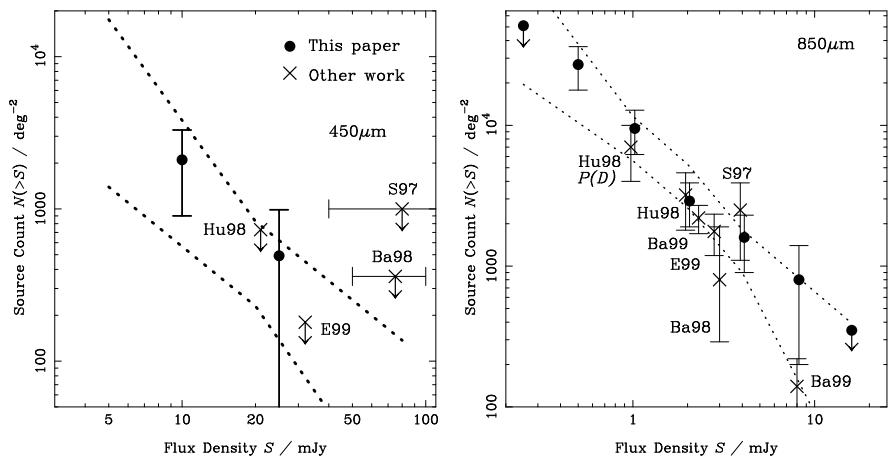

\begin{center}
\plotfiddle{blaina1_1.ps}{5.1cm}{-90}{38}{38}{-230}{190}
\end{center}
\begin{center}
\plotfiddle{blaina1_2.ps}{4.3cm}{-90}{38}{38}{-60}{337}
\end{center}
\vskip -5cm
\caption{ 
Left: the 450-$\mu$m counts of galaxies. The direct and Monte-Carlo 
counts derived here are shown by the solid points and dotted lines 
respectively. Right: the 850-$\mu$m counts of galaxies including 
the updated SCUBA lens survey counts. To avoid complicating the 
figure the direct counts obtained by Blain et al.\ (1999; see Table\,1) 
are not shown. The associated Monte-Carlo results are shown by 
the dotted lines. Ba98/Ba99 -- Barger et al.\ (1998, 1999a); 
E99 -- Eales et al.\ (1999); Hu98 -- Hughes et al.\ (1998); S97 -- 
Smail et al.\ (1997). 
}
\end{figure}

%
%


\acknowledgments

We are glad to be able to help acknowledge Hy's 65 years. AWB 
thanks the conference organizers for their hospitality and for 
support during the meeting, and also Jon Lacey and Jackie Davidson 
for good food before and after. The JCMT is operated by the 
Observatories on behalf of the UK PPARC, the Netherlands SRON and 
the Canadian NRC.


%
%

%

\end{document}